\documentclass[prl,amsmath,amssymb,
 reprint,%
]{revtex4-1}

\usepackage{graphicx}	
\usepackage{dcolumn}	
\usepackage{bm}		
\usepackage{pstricks}
\usepackage{graphicx}
\usepackage{multimedia}
\usepackage{textpos} 
\usepackage{lmodern}
\usepackage{amsmath,amssymb,fancyhdr,epsfig,natbib,color}
\usepackage{pgf}

\newcommand{\ud}{\mathrm{d}}
\newcommand{\uD}{\mathrm{D}}

\newcommand{\vv}[1]{\bm #1}

\usepackage{color}
\definecolor{myb}{rgb}{0.03,0.03,1}
\definecolor{myg}{rgb}{0.1647,0.72,0.1647}
\definecolor{myr}{rgb}{0.84,0,0.078}
\definecolor{mym}{rgb}{0.84,0,0.78}
\definecolor{myy}{rgb}{0.84,0.84,0}
\definecolor{mybr}{rgb}{0.5,0.,0}
\definecolor{mygr}{rgb}{0.5,0.5,0.5}
\definecolor{myc}{rgb}{0,0.88,0.98}
\definecolor{mya}{rgb}{0,0.6627,1}
\definecolor{myo}{rgb}{1,0.61,0}

\usepackage{graphicx}
\usepackage{dcolumn}
\usepackage{bm}
 \usepackage{relsize}
 \usepackage{wasysym}

\begin{document}
\preprint{AIP/123-QED}


\title{The critical impact speed for the splash of a drop}

\author{Guillaume Riboux \& José Manuel Gordillo}
\affiliation{\'Area de Mec\'anica de Fluidos, Departamento de Ingener\'ia
Aeroespacial y Mec\'anica de Fluidos, Universidad de Sevilla,
Avenida de los Descubrimientos s/n 41092, Sevilla, Spain.}

\date{\today}

\date{\today}

\begin{abstract}
Making use of experimental and theoretical considerations, in this Letter we deduce a criterion to determine the critical velocity for which a drop impacting a smooth dry surface, either spreads over the substrate or disintegrates into smaller droplets. The derived equation, which expresses the splash threshold velocity as a function of the material properties of the two fluids involved, the drop radius and the mean free path of the molecules composing the surrounding gaseous atmosphere, has been thoroughly validated experimentally at normal atmospheric conditions using eight different liquids, with viscosities ranging from $\mu=3\times 10^{-4}$ to $\mu=10^{-2}$ Pa$\cdot$s and interfacial tension coefficients varying between $\sigma=17$ and $\sigma=72$ mN$\cdot$m$^{-1}$. Our predictions are also in fair agreement with the measured critical speed of drops impacting in different gases at reduced pressures given by Xu \emph{et al}\cite{Nagel}.
%
\end{abstract}
\pacs{Valid PACS appear here}
\keywords{Suggested keywords}

\maketitle

The collision of a drop against a solid surface is ubiquitous in Nature and is present in a myriad of technological and scientific fields comprising ink-jet printing, combustion or surface coating \cite{Nagel,Bergeron,PRLLohseMaxbub}. Given the physical properties of both the liquid and the gas, the atmospheric pressure \cite{Nagel}, the size of the drop and the physicochemical properties of the substrate \cite{Duez}, experience reveals that there exists a critical impact velocity below which the liquid simply spreads over the surface and above which the original liquid volume fragments into tiny droplets violently ejected outwards, creating what is known as a splash (see Fig. 1). In spite of the number of advances on the subject \cite{Mundo,Nagel,PRLBrenner,Duchemin,Nagelrough,Rubinstein,JosserandZaleski,Stonesplash,Rioboo,Yarin,Palacios}, a precise description of the critical conditions leading to drop splashing, is still lacking \cite{Rein,AnnRevSnoeijer}. Indeed, the well known equation deduced twenty years ago by Mundo \emph{et al} \cite{Mundo,Slingshot,Stonesplash}, as well as many empirical correlations \cite{Palacios}, provide expressions for the critical velocity which depend only on the material properties of the liquid, and do not take into account that splashing is largely affected by the gaseous atmosphere \cite{Nagel}. In this Letter, we provide a theoretical framework which is consistent with our own experimental data and also with \emph{all} previous findings.


\begin{figure}[ht!]
\begin{center}
\includegraphics[width=0.42\textwidth]{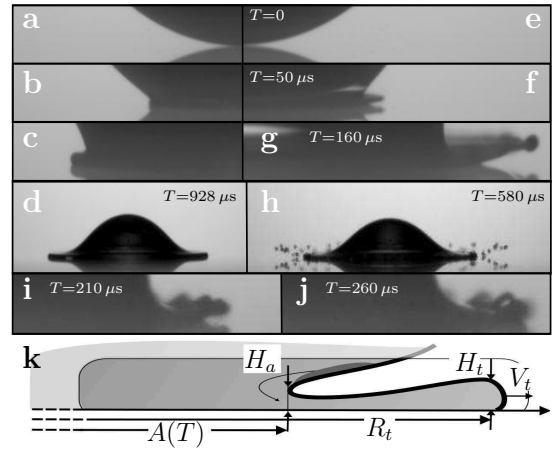}
\caption{(a)--(h): Sequence of events after the impact of an ethanol droplet of radius $R=$1.04 mm for two different impact velocities, $V$=1.29\,m$\,$s$^{-1}$ (1(a)--(d), left) and $V$=2.28\,m$\,$s$^{-1}$ (1(e)--(h), right). The drop simply spreads over the substrate for the smaller value of $V$ but breaks into tiny drops, violently ejected outwards, for the larger impact velocity. Note from images 1(f)--(g) that the lamella dewets the substrate before creating the splash depicted in Fig. 1(h). The sequence of events represented in Figs. 1(i)--(j), for which $V$=2.01\,m$\,$s$^{-1}$, illustrates the existence of an intermediate range of impact velocities for which the lamella firstly dewets the solid to contact it again as a consequence of the radial growth of the edge of the lamella. The splash threshold velocity corresponding to these experiments is $V$=2.19\,m$\,$s$^{-1}$. The times in 1(a)--(c) are identical to those corresponding to images 1(e)--(g). The sketch in Fig. 1(k) illustrates the definition of the main variables used along the paper.\label{fig1}}
\end{center}
\vspace*{-0.7cm}
\end{figure}

To elucidate the precise conditions under which a drop hitting a solid surface splashes or not, we perform experiments with millimetric drops of radii $R$ formed quasi-statically at normal atmospheric conditions. Eight different liquids are slowly injected through hypodermic needles of different diameters. Drops generated in this way are spherical and fall under the action of gravity onto a dry glass slide, with a composition such that the liquids, whose physical properties are listed in Table 1 of the supplementary material section, partially wet the substrate with a static contact angle $\sim 20^o$. The impact speed $V$ is varied by fixing the vertical distance between the exit of the needles and the impactor. To simultaneously record the impact process from the side with two different optical magnifications and acquisition rates, two high speed cameras focusing the impact region are placed perpendicularly to each other.

Figure 1 shows the detailed sequence of events recorded from the instant $T$=0 at which the drop first contacts the solid. These images reveal that, initially, the drop deforms axisymmetrically, with $A(T)$ the radius of the circular wetted area [Figs. 1(a)--(b)] and that an air bubble is entrapped at the center of the drop  \cite{Thorodd,PRLLohseMaxbub}; however, the presence of this tiny bubble does not affect the splash process. Figure 1(c) illustrates that for $T\geq T_e$, a thin sheet of liquid starts to be expelled from the radial position where the drop contacts the solid, i.e. $A(T_e)$, with $T_e$ the ejection time. Of special relevance for the purposes of this study is to observe the change of trajectory experienced by the edge of the sheet as the impact velocity increases. Indeed, for the smallest values of $V$ [Figs. 1(a)--(d)], the lamella spreads tangentially along the solid but, for a range of larger impact velocities, the liquid initially dewets the substrate and contacts the substrate again [Figs. 1(i)--(j)]. For even higher values of $V$, the front of the lamella dewets the solid [Figs. 1(f)--(g)] and drops are finally ejected radially outwards [Figs. 1(g)--(h)] in a way similar to the experiments reported in \cite{Thoroddsen2,VillermauxBossa,JFMSplash13}.
Therefore, the analysis of the images in Fig. \ref{fig1} reveals that, for a splash of the type illustrated in Figs. \ref{fig1}(e)--(h) to take place, two conditions need to be fulfilled simultaneously: the liquid must dewet the solid and the vertical velocity imparted to the front part of the lamella needs to be large enough to avoid the liquid to contact the solid again. To obtain a splash criterion, it is essential to observe from Figs. 1(i)--(j) and the movies in the supplementary material section, that the rewetting is a consequence of the radial growth of the rim thickness, $H_t$, caused by capillary retraction. Along the Letter, times, velocities and pressures are made dimensionless using $R$, $V$, $R/V$, $\rho\,V^2$ as characteristic length, velocity, time and pressure, with $\rho$ the liquid density and lower-case letters denoting dimensionless variables; the subscript $g$ will be used to denote gas quantities. Next, the instant $T_e$ at which the lamella is ejected, as well as its initial height and velocity, $H_t$ and $V_t$ respectively (see figure 1k), will be calculated.

\begin{figure}
\begin{center}
\includegraphics[width=0.35\textwidth]{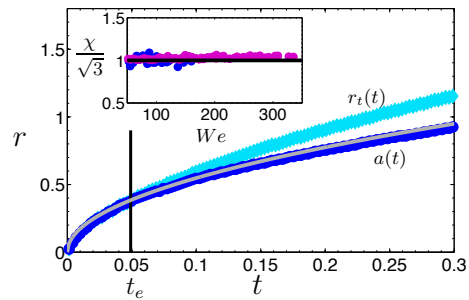}
\caption{Experimental radius of the wetted area compared with $a$=$\sqrt{3\,t}$ (solid line) for $We$=98, $Re$=3462 (water drop). The radial position of the ejecta sheet, $r_t$, is also represented for times $t\,$$\geqslant$$\,t_e$, with $t_e$ the ejection time. The initial velocity of the ejected sheet coincides with that of the radius of the wetted area at $t$=$t_e$ since both $r_t(t)$ and $a(t)$ are tangent to each other at the ejection time. The inset represents the ratio $\chi/\sqrt{3}$$\simeq$1, where $\chi$ is the coefficient obtained from the best fit of a function of the type $a$=$\chi\sqrt{t}$ to the experimentally measured radius $a(t)$, for a large range of Weber numbers and two different liquids: water ({\fontsize{12}{24}\selectfont\color{myb}{$\bullet$}}: $\mu$$\simeq$0.9\,cP, $\sigma$$\simeq$67.5\,mN\,m$^{-1}$) and a silicon oil ({\fontsize{12}{24}\selectfont\color{mym}{$\bullet$}}) of viscosity $\mu$=10\,cP and surface tension $\sigma$$\simeq$19.5\,mN\,m$^{-1}$. The ratio $\chi/\sqrt{3}\simeq 1$ also for the rest of the fluids investigated.\label{fig2}}
\end{center}
\vspace*{-0.7cm}
\end{figure}

Splashing occurs when the values of both the Weber and Reynolds numbers are such that $We$=$\rho\,V^2\,R/\sigma$$\gg$1, $Re$=$\rho\,V\,R/\mu$$\gg$1, with the dimensionless numbers $We$ and $Re$ measuring the relative importance of inertial and surface tension stresses ($We$) and inertial and viscous stresses ($Re$). Consequently, during the characteristic impact time, $R/V$, viscous effects are confined to thin boundary layers of typical width $\sim R\,Re^{-1/2}$$\ll$$R$ \cite{Schlichting} a fact suggesting that the use of potential flow theory \cite{Wagner,AnnrevKorobkin}, which neglects liquid viscosity, is appropriate to describe the liquid flow at the scale of the liquid drop. For the sake of clarity, we report here only the main results of the analysis, being further details provided in the Supplementary Material Section, where using Wagner's theory \cite{Wagner} we deduce that the radius of the wetted region evolves in time as $a(t)$=$\sqrt{3t}$ \cite{Mongruel}. Potential flow theory also predicts that, as a consequence of the sudden inertial deceleration of the liquid when it hits the wall, a flux of momentum is directed tangentially along the substrate \cite{JFMSplash13}, giving rise to the ejection of a fast liquid sheet, like the one depicted in Figs. 1(c) and 1(f)--(g). The application of the Euler--Bernoulli equation at the drop's interface, where the pressure remains constant, in a frame of reference moving at a velocity $\dot{a}$ (see figure 1k), yields that fluid particles are ejected from $a(t)$ at a speed relative to that of the ground given by $v_a$=$2\dot{a}$=$\sqrt{3/t}$. Moreover, since the flux of tangential momentum per unit length is $\propto \rho\,V^2\,R\,a(t)$ \cite{JFMSplash13},\footnote{In \cite{JFMSplash13} it is shown that the ejected flux per unit length of tangential momentum is proportional to $\rho V^2 R s$, with $R s$ the radius of the impacting region. In the present case, $s=a(t)$}, we thus conclude that the height $h_a$ of the lamella at the intersection with the spreading drop, i.e. at the radial position $r$=$a(t)$, is $\rho\,V^2\,R\,\dot{a}^2\,h_a\propto\rho\,V^2\,R\,a\Rightarrow h_a\propto a/\dot{a}^2\propto t^{3/2}$, with dots denoting time derivatives \cite{Howison,Oliver,SKorobkin_JFS}.

Figure \ref{fig2} shows that the measured radius of the wetted area perfectly matches $a$=$\sqrt{3t}$ for all the different impact events and fluids considered, a result that fully validates our potential flow calculation. However, while our experimental evidence indicates that the ejecta sheet is only produced for $t\,$$\geq$$\,t_e$, the potential flow approach predicts the generation of a lamella for $t\,$$\geq$0 of vanishingly small thickness $h_a$$\propto$$\,t^{3/2}$ with fluid velocity diverging as $v_a$$\propto$$\,t^{-1/2}$.

To understand the differences between potential flow results and observations note first that, in analogy with the cases of bubbles bursting at a free interface \cite{MacIntyre} and Worthington jets \cite{PRL09,JFM10}, the fluid feeding the lamella comes from a region where shear stresses are negligible, namely, a very narrow boundary straddling the drop's interface (dark shaded region in panel 1k), \emph{and not} from the boundary layer growing from the stagnation point located at the axis of symmetry.

\begin{figure}
\begin{center}
\hspace*{-4mm}
\includegraphics[width=0.4\textwidth]{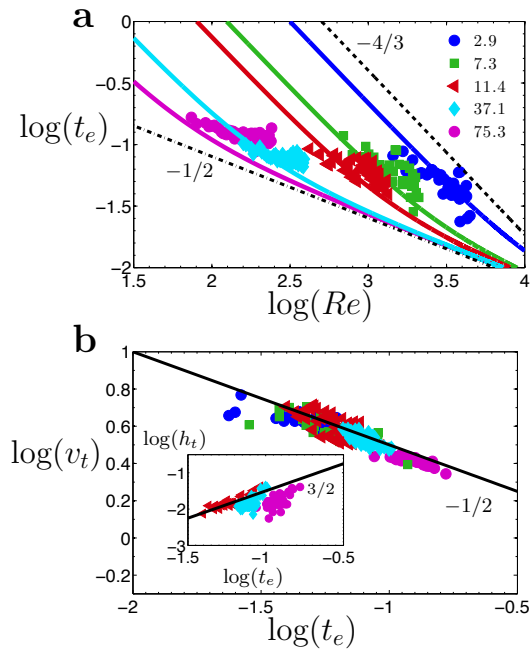}
\caption{(a) Comparison between the experimentally measured value of the ejection time, and the one calculated solving equation (\ref{te}). Each of the solid lines represent the different theoretical results for different Ohnesorge numbers. The values of 1000$\times$$Oh$, are represented in the legend. The experimentally determined ejection time is very well approximated by the solution of equation (\ref{te}) and clearly follow the low and high Ohnesorge number limits reflected by equation (\ref{te}), $t_e\propto Re^{-4/3}\,Oh^{-4/3}$ and $t_e\propto Re^{-1/2}$. (b) The measured experimental data follow our prediction $v_t\propto t_e^{-1/2}$ and $h_t\propto t_e^{3/2}$. We find that $h_t\simeq 2.8\,h_a$, with $h_a=\sqrt{12}/\pi\,t^{3/2}_e$ deduced in the supplementary material section\label{fig3}}
\end{center}
\vspace*{-0.7cm}
\end{figure}

Fast fluid particles entering the liquid sheet are rapidly decelerated within the lamella due the combined action of both the viscous shear stresses diffusing from the wall and capillary pressure. The characteristic thickness $\delta$ of the region affected by viscous stresses at a distance $\sim$$h_a$$\ll$1 downstream the jet root (see Fig. 1k), which is the region where the jet meets the drop, is $\delta/h_a$$\sim$1 (see Supplementary Materials Section). The deceleration, provokes the fluid to accumulate at the edge of the liquid sheet and, consequently, $h_t>h_a$ (see Fig. 1k).

To determine $t_e$, note from Fig. \ref{fig2} that, although the velocities of fluid particles entering the jet are $v_a=2\dot{a}>\dot{a}$, both $a(t)$ and $r_t$ are tangent to each other at the instant of ejection, namely, $v_t=\dot{a}$ at $t=t_e$. Thus, since the lamella can only be ejected if its tip advances faster than the radius of the wetted area, the condition for sheet ejection is $\uD v/\uD t\,=-\partial\,p/\partial x+Re^{-1} \nabla^2 v\geq$$\ddot{a}$ at $t$=$t_e$. Here, $\uD v/\uD t\,$$<$0 is the dimensionless acceleration of the material points in the sheet given by the momentum equation,
$p$ denotes pressure and $x$ measures the distance from the jet root. To determine $t_e$, since $\delta$$\sim$$h_t$,  $\nabla^2 v$$\sim $$\dot{a}/h^2_t\propto t_e^{-1/2}/h^2_t$; moreover, the increment of pressure experienced by fluid particles flowing into the edge of the lamella is the capillary pressure $We^{-1}/h_t$ and thus, $\partial p/\partial x$$\sim$$Re^{-2}Oh^{-2}/h^2_t$ with $\Delta x$$\sim$$h_t$ and $Oh=\mu/\sqrt{\rho\,R\,\sigma}=\sqrt{We}/Re$ the Ohnesorge number. Therefore, the critical condition for sheet ejection $\uD v/\uD t=\ddot{a}\propto t_e^{-3/2}$ is
\begin{equation}
c_1\,Re^{-1}\,t_e^{-1/2}+Re^{-2}\,Oh^{-2}=\ddot{a}\,h^2_t=c^2 t_e^{3/2}\, ,\label{te}
\end{equation}
where we set $c_1=\sqrt{3}/2$ for simplicity and $c=1.1$ accounts for the proportionality constant in $h_t$$\propto$$h_a$$\propto$$t_e^{3/2}$. Figure \ref{fig3}a, illustrates that equation (\ref{te}) very well approximates the experimental results. Experiments also validate the high--$Oh$ and low--$Oh$ limits of Eq. (\ref{te}), respectively given by  $Re^{-1}\,t_e^{-1/2}\propto t^{3/2}\Rightarrow t_e\propto Re^{-1/2}$ and $Re^{-2} Oh^{-2}\propto t_e^{3/2}\Rightarrow t_e\propto Re^{-4/3} Oh^{-4/3}$. In addition, Fig. \ref{fig3}b reveals that, $v_t\propto t_e^{-1/2}$ and $h_t\propto t_e^{3/2}$, providing further support to our theory.

\begin{figure}[t!]
\begin{center}
\includegraphics[width=0.4\textwidth]{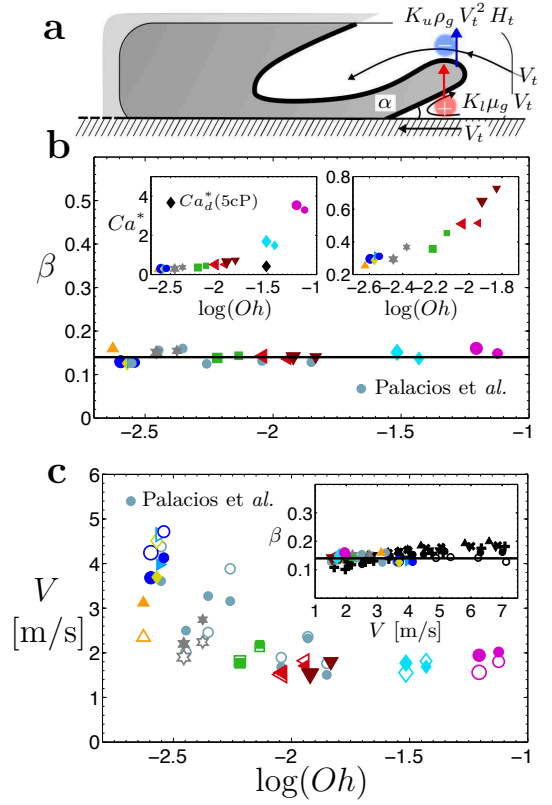}
\caption{(a) The total lift force arises as the addition of the force in the wedge of angle $\alpha$ and the force at the top part of the lamella. The plus/minus signs indicate the regions where the gauge pressure is positive/negative respectively. (b) Values of the function $\beta$ calculated through Eq. (\ref{condicionbeta}) for the different liquids and drop diameters investigated. Experimental data from \cite{Palacios} have also been included. The inset represents the values of $Ca^*$=$\left(\mu V/\sigma\right)\sqrt{3}/2 t^{-1/2}_{e,crit}$, with $t_{e,crit}$ the solution of equation (\ref{te}) for the experimental splash threshold velocity given in Fig. 4(c). (c) The open symbols represent the splash threshold velocities predicted by the solution of equations (\ref{te}) and (\ref{condicionbeta}) with $\beta \simeq$ 0.14. The inset shows that the splash threshold velocity $V$ corresponding to the experiments in Xu \emph{et al} \cite{Nagel} is also characterized by $\beta\simeq 0.14$. Our own data, as well as those in \cite{Palacios} are also included in the inset. The symbols in black have the same meaning as in Xu \emph{et al} \cite{Nagel}.\label{fig4}}
\end{center}
\vspace*{-0.7cm}
\end{figure}

Now, following the ideas in \cite{Rein,Duez}, we represent in the inset of Fig. \ref{fig4}(b) the capillary number, $Ca^*$, defined using the value of $V_t$ at the splash transition. In Duez \emph{et al}\cite{Duez}, $Ca^*$ is constant for the case of wetting surfaces and low viscosity liquids because, in their case, the splash and dewetting transitions coincide. However, since dewetting is a necessary but not sufficient condition for drop splashing in our case (see Figs. 1 i--j), we find that i) $Ca^*$ varies appreciably with the liquid properties and ii) $Ca^*$ is larger than the critical capillary number $Ca^*_{d}$ above which the liquid dewets the substrate. Indeed, the type of lubrication equations in \cite{Marchand} representing a static force balance \emph{in the direction tangent to the wall}, are integrated to determine $Ca^*_d$ (see the Supplementary Material for details) and the result, particularized for $\mu$=5\,cP and illustrated in the inset of Fig. \ref{fig4}(b), reveals that $Ca^*$ is well above $Ca^*_{d}$.

Thus, to determine the critical speed of an impacting drop, we note first that splashing occurs as a consequence of the \emph{vertical} lift force $\mathbf{\ell}$, imparted by the gas on the edge of the liquid sheet. The lift force results from the addition of two contributions: the lubrication force $\sim K_l\mu_g V_t$ and the suction force $\sim K_u \rho_g V^2_t\,H_t$. The former is exerted at the wedge formed between the substrate and the edge of the lamella and the latter, at the top part of it (see Fig. \ref{fig4}a). In the Supplementary Material Section we show that $K_l$$\simeq$ $-(6/\tan^2(\alpha))\left[\ln\left(19.2\lambda/H_t\right)-\ln\left(1+19.2\lambda/H_t\right)\right]$, with $\lambda$ the mean free path of gas molecules, $\alpha$ the wedge angle which, for the case of partially wetting solids considered here does not seem to significatively depend on liquid viscosity\footnote{The value of $\alpha$ should be dependent on the wetting properties of the substrate but, in the case of our experiments, all the liquids partially wet the substrate with a similar contact angle $\sim 20^o$} and $K_u$$\simeq $$0.3$. Since the local gas Reynolds number based on $H_t$ and $V_t$ is $\sim O(10)$, both the viscous and inertial contributions to the lift force need to be taken into account (see the Supplementary Material for details). Therefore, the vertical force balance per unit length, when applied at the edge of the lamella, reads $\rho\,H^2_t\,\dot{V}_v\sim \mathbf{\ell}$
with $V_v$ the vertical velocity, whose characteristic value at the instant when the liquid front has raised a distance $\sim$$H_t$ above the substrate is given by
\begin{equation}
V_v\sim \sqrt{\mathbf{\ell}/\left(\rho\,H_t\right)}\quad\mathrm{where}\quad \mathbf{\ell}=K_l\mu_g\,V_t+K_u\rho_g\,V^2_t\,H_t\, .\label{vv}
\end{equation}

For the edge of the sheet not to contact the solid again (see Fig. \ref{fig1}i-j), $V_v$ needs to be larger than $\dot{R}_c$. Here, $R_c$ is the rim radius of curvature which, once the liquid dewets the substrate, grows in time as a consequence of capillary retraction. Naming $V_{r}$ the well known Taylor--Culick velocity given by the momentum balance $2\sigma=\rho\,V^2_{r}\,H_t$,
\begin{equation}
V_{r}=\sqrt{2\sigma/\rho\,H_t} \label{VTC}
\end{equation}
and inserting this result into the mass balance $\ud\,R^2_c/\ud T$$\sim$$V_{r}\,H_t$, one readily obtains that, at the instant when the liquid separates from the substrate, for which $R_c\simeq H_t/2$, $\dot{R}_c\sim V_{r}$. Hence, the splash threshold condition reads $\beta=V_v/V_{r}\propto \left(\mathbf{\ell}/\sigma\right)^{1/2}\sim O(1)$, with $V_v$ and $V_{r}$ respectively given in Eqs. (\ref{vv})--(\ref{VTC}). Using the expression for the lift force $\mathbf{\ell}$ in Eq. (\ref{vv}), the splash condition can thus be expressed as
\begin{equation}
\beta=\left(\frac{K_l\mu_g\,V_t+K_u\rho_g\,V^2_t\,H_t}{\sigma}\right)^{1/2}\sim O(1)\, .\label{condicionbeta}
\end{equation}
To check the validity of Eq. (\ref{condicionbeta}), Eq. (\ref{te}) is used first to calculate the ejection time $t_{e,crit}$ corresponding to the values of $We$, $Re$, $\rho/\rho_g$ and $\mu/\mu_g$ for which the splash transition is experimentally observed. Once $t_{e,crit}$ is known, $V_t=\sqrt{3}/2 V\,t^{-1/2}_{e,crit}$, $H_t=2.8\,R\,\sqrt{12}/\pi\,t^{3/2}_{e,crit}$ and $\beta$ is determined through Eq. (\ref{condicionbeta}). Fig. \ref{fig4}(b) demonstrates that the splash threshold is characterized by a nearly constant value of $\beta$, independent of the type of liquid considered, as predicted by Eq. (\ref{condicionbeta}). The open symbols in Fig. \ref{fig4}(c), representing the splash threshold velocities calculated solving equations (\ref{te}) and (\ref{condicionbeta}) for $\beta\simeq 0.14$, are fairly close to those measured experimentally, a fact further supporting our theory. Interestingly enough, the inset in Fig. \ref{fig4}(c) shows that the splash threshold corresponding to \emph{all the experimental data} in Xu \emph{et al}\cite{Nagel}, where the critical speed of drops of different liquids falling within several gases and different pressures is investigated, is also characterized by $\beta\simeq 0.14$.

Since the lift force $\mathbf{\ell}$ is dominated by the term $K_l\mu_g\,V_t$ (see the Supplementary Material for details) and $K_l$ is approximately constant because of its logarithmic dependence on the physical parameters, the splash criterion (\ref{condicionbeta}) \emph{at normal atmospheric conditions} can be \emph{approximated} by $\left(\mu_g\,V/\sigma\right)\,t_{e,crit}^{-1/2}\propto  C$, with $t_{e,crit}$ the solution of equation (\ref{te}) and $C$ a constant. Due to the fact that $t_{e,crit}\propto Re^{-4/3}\,Oh^{-4/3}$ (see Fig. \ref{fig3}a) in the low--$Oh$ limit, the previous approximate criterion results in $Oh_g\,Oh^{5/3}\,Re^{5/3}\equiv \left(Re\,Oh_g^{8/5}\right)^{5/3}(\mu/\mu_g)^{5/3}\propto C'$, where $Oh_g=\mu_g/\sqrt{\rho R\sigma}$. In the high--$Oh$ limit, $t_{e,crit}\propto Re^{-1/2}$ (see Fig. \ref{fig3}a) and, in this case, the approximate splash criterion is $Oh_g\,Oh\,Re^{5/4}\equiv \left(Re\,Oh_g^{8/5}\right)^{5/4}\left(\mu/\mu_g\right)\propto C''$. These power law expressions are experimentally validated in the Supplementary Material Section, giving a physical explanation to the well-known correlation by Mundo \emph{et al} \cite{Mundo,Rein,Palacios} and to the interesting finding in \cite{Deegan} that the splash threshold condition cannot be solely characterized in terms of $Oh$ and $Re$. Let us emphasize that these power law expressions are simply approximations to our theory, expressed by equations (1) and (4).

To conclude, we have deduced a criterion expressing the splash threshold velocity of a drop impacting on a smooth, dry surface as a function of the liquid density and viscosity, the millimetric drop radius, the gas density and viscosity and the nanometric mean free path of gas molecules.

The authors wish to express their most sincere gratitude to Professor Alexander Korobkin for useful suggestions, discussions, for providing them with many relevant references on the subject and for his kind invitation to JMG to participate in the seminar \emph{Mathematics of Splashing}, Edinburgh, June 2013. Useful comments by Alejandro Sevilla, Javier Rodr\'iguez-Rodr\'iguez and Devaraj van der Meer are also very much acknowledged. We are also grateful to Alonso Fernández for providing us with the numerical values of $K_u$ as a function of the local gas Reynolds number. This work has been supported by the Spanish MINECO under Project DPI2011-28356-C03-01, which has been partly financed through European funds.

%

\clearpage
\newpage

\section*{Supplementary Material}


\subsection*{Experiments}

Since the liquids with physical properties given in table 1 are injected quasi-statically, the radius $R$ of the drops generated are approximately given by $R/D_c\propto \left(\ell_\sigma/D_c\right)^{2/3}$, with $D_c$ the diameter of the injection tube, $\ell_\sigma=\left(\rho\,g/\sigma\right)^{-1/2}$ the capillary length, $\rho$ the liquid density, $g$ the acceleration of gravity and $\sigma$ the interfacial tension coefficient. The acquisition rate of the camera used to describe the overall impact process varied between 17241 and 29197 frames per second and the resolution in microns of the captured images varied between 16.60 to 31.91 microns/pixel. The high speed camera used to record the impact details was operated between $10^5$ and 641509 frames per second, providing spatial resolutions ranging from 4.23 to 14.00 microns/pixel.

\begin{figure}
\begin{center}
\includegraphics[width=0.48\textwidth]{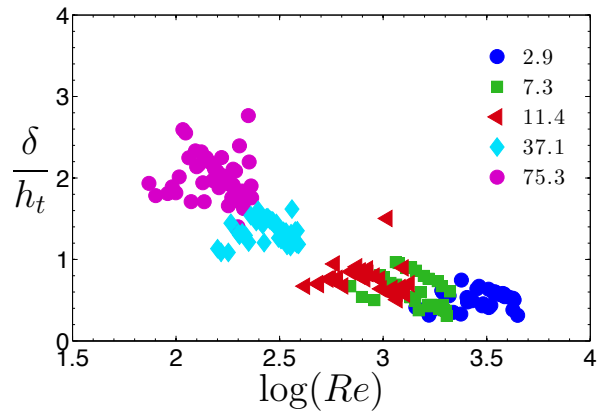}
\caption{\label{figA4} Experimentally obtained values of the ratio $\delta/h_t$$\sim$$ 5\left(Re\,t_e\right)^{-1/2}$ for some of the fluids used in this study. The prefactor of $5$ used in the estimation of $\delta$ is motivated by the factor in the thickness of boundary layers growing with zero pressure gradient\cite{Schlichting}. The values in the legend represent $1000$$\times$$Oh$.}
\end{center}
\end{figure}

The characteristic thickness $\delta\sim \sqrt{\nu\,t}$ of the region affected by viscous stresses at a distance $\sim$$h_a$$\ll$1 downstream the jet root (see Fig. 1k), which is the region where the jet meets the drop, with $t\sim h_a/v_a$ is
\begin{equation}
\frac{\delta}{h_a}\propto Re^{-1/2} \left(h_a\,v_a\right)^{-1/2}\sim\left(Re\,t\right)^{-1/2}\, .\label{delta_a}
\end{equation}
For all our experimental data, $\delta/h_a$$\sim$1 (see Fig. \ref{figA4}), and thus the lamella is decelerated by viscous shear stresses \cite{JosserandZaleski} at a tiny distance $\sim h_a$ from the jet root and also by the capillary pressure. The deceleration also provokes the fluid to accumulate at the edge of the liquid sheet and, consequently, $h_t>h_a$ (see Fig. 1k).
The analysis of the experimental data depicted in Fig. \ref{figA4} reveals that the characteristic boundary layer thickness based on the experimentally measured values of the velocity and height of the edge of the lamella, is always close to the thickness of the lamella.

\begin{table}
\caption{Physical properties of the different fluids used, drop radius and the corresponding Ohnesorge numbers.}
\begin{center}
\begin{ruledtabular}
\begin{tabular}{cccccccc}
 & & $\rho$ & $\sigma$ & $\mu$ & $R$ & $\ell_\sigma$  & $Oh\times10^3$\\
 & & (kg/m$^{3}$) &  (mN/m) &  (cP) &  (mm) & (mm) & (-)\\
 ($a$) & {\fontsize{9}{18}\selectfont\color{myo}{$\blacktriangle$}} & 789 & 24.0 & 0.3 & 1.03 & 1.76 & 2.4\\
 ($b$) & {\fontsize{15}{27}\selectfont\color{myb}{$\bullet$}} & 1000 & 71.8 & 0.95 & 1.96 & 2.71 & 2.5\\
    & {\fontsize{9}{18}\selectfont\color{myy}{$\blacklozenge$}} & 1000 & 71.8 & 0.95 & 1.74 & 2.71 & 2.7\\
   & {\fontsize{9}{18}\selectfont\color{myc}{$\blacktriangleright$}} & 1000 & 71.8 & 0.95 & 1.63 & 2.71 & 2.8\\
   & {\fontsize{12}{24}\selectfont\color{myb}{$\bullet$}} & 1000 & 67.5 & 0.9 & 1.45 & 2.62 & 2.9\\
 ($c$) & {\fontsize{10}{19}\selectfont\color{mygr}{$\bigstar$}} & 791 & 23.5 & 0.6 & 1.53 & 1.74 & 3.5\\
    & {\fontsize{8}{17}\selectfont\color{mygr}{$\bigstar$}} & 791 & 23.5 & 0.6 & 1.05 & 1.74 & 4.2\\
 ($d$) & {\fontsize{9}{17}\selectfont\color{myg}{$\blacksquare$}} & 789 & 22.6 & 1.0 & 1.53 & 1.71 & 6.1\\
  & {\fontsize{6}{14}\selectfont\color{myg}{$\blacksquare$}} & 789 & 22.6 & 1.0 & 1.04 & 1.71 & 7.3\\
 ($e$) & {\fontsize{11}{20}\selectfont\color{myr}{$\blacktriangleleft$}} & 854 & 17.2 & 1.3 & 1.34 & 1.43 & 9.1\\
  & {\fontsize{8}{17}\selectfont\color{myr}{$\blacktriangleleft$}} & 854 & 17.2 & 1.3 & 0.86 & 1.43 & 11.4\\
 ($f$) & {\fontsize{11}{20}\selectfont\color{mybr}{$\blacktriangledown$}} & 875 & 17.8 & 1.7 & 1.37 & 1.44 & 12.0\\
   & {\fontsize{8}{17}\selectfont\color{mybr}{$\blacktriangledown$}} & 875 & 17.8 & 1.7 & 0.92 & 1.44 & 14.7\\
 ($g$) & {\fontsize{11}{20}\selectfont\color{myc}{$\blacklozenge$}} & 913 & 18.6 & 4.6 & 1.32 & 1.44 & 30.5\\
   & {\fontsize{8}{17}\selectfont\color{myc}{$\blacklozenge$}} & 913 & 18.6 & 4.6 & 0.89 & 1.44 & 37.1\\
 ($h$) & {\fontsize{15}{27}\selectfont\color{mym}{$\bullet$}} & 1000 & 19.5 & 10.0 & 1.32 & 1.41 & 62.2\\
  & {\fontsize{12}{24}\selectfont\color{mym}{$\bullet$}} & 1000 & 19.5 & 10.0 & 0.90 & 1.41 & 75.3\\
\end{tabular}
\end{ruledtabular}
($a$) Acetone, ($b$) Water, ($c$) Methanol, ($d$) Ethanol, ($e$) Decamethyltetrasiloxane, ($f$) Dodecamethylpentasiloxane, ($g$) Poly(Dimethylsiloxane) and ($h$) 10 cP Silicone Oil. $Oh=\sqrt{We}/Re=\mu/\sqrt{\rho\,R\sigma}$.
\label{tab1}
\end{center}
\end{table}

\subsection*{Determination of a(t)}

Due to the fact that $Re=\rho\,V\,R/\mu\gg 1$, the velocity field sufficiently far from the wall can be expressed in terms of a velocity potential $\vv{u}$=$\nabla\,\phi(r,z,t)$ which, by virtue of the continuity equation, $\nabla\cdot\vv{u}$=0, verifies the Laplace equation $\nabla^2\phi$=0. To satisfy the impermeability boundary condition $\vv{u}\cdot\mathbf{k}$=0 at the circular region of radius $a(t)$ formed by the intersection of the drop with the substrate [see Fig. \ref{Volcontrol}], the flow field within the drop can be expressed as the addition of two velocity fields: the one associated with the impact velocity ($-\mathbf{k}$) plus $\vv{u'}$=$\nabla\phi'$, which verifies the Laplace equation subjected to the the condition at $z$=0, $\mathbf{k}\cdot\vv{u'}$=1 for $r$$\leq$$a(t)$ and to the Euler--Bernoulli equation at the drop interface $r\,$$>$$\,a(t)$, $z$=$z_d(r,t)$, $\partial\phi'/\partial\,t+u^2/2\simeq 1/2$ with $z_d$ the vertical height of the drop interface with respect to the wall. In the previous expression, we have taken into account the fact that, since $We\gg 1$, the contribution of the capillary pressure $\sigma/R$ has been neglected with respect to $\rho\,V^2$. Now note that the boundary condition at the free interface can be further simplified due to the fact that, during the initial instants, the radius of the wetted area verifies the condition $a(t\ll 1)\ll 1$ and thus, since at $t$=0, $z_d$=$r^2/2$, the position of the free interface can be approximated for $t\ll 1$ as $z_d\simeq 0$ with errors $\sim O(a^2)\sim O(t)$. Moreover, for $t\ll 1$, the Euler--Bernoulli equation simplifies to $\partial\phi'/\partial\,t\simeq 0$ since the local acceleration term dominates over the convective one \cite{AnnrevKorobkin,JFMSplash13}, implying that the Euler--Bernoulli equation at the free interface, which is approximately located at $z$=$0$, simplifies to $\phi'$=0 with errors of the order of $O(t)\ll 1$ due to the fact that, at $t=0$, $\phi'=0$ at the free interface. Consequently, the analytical solution of $\nabla^2\phi'$=0 subjected to the boundary conditions at $z$=0, $\mathbf{k}\cdot\mathbf{u'}$=1 for $r\leq a(t)$, $\phi'$=0 for $r\,$$>$$\,a(t)$ and to $\nabla\phi'$=0 for $z\rightarrow \infty$ leads to the following expression for the normal velocity at $z$=0 and $r\,$$>$$\,a(t)$ (see \cite{Lamb}):
\begin{equation}
\mathbf{k}\cdot\vv{u}=-1-\frac{2}{\pi}\left[\frac{a}{\sqrt{r^2-a^2}}-\arcsin\left(\frac{a}{r}\right)\right]\, .\label{Eq2}
\end{equation}

Now, the equation that determines $a(t)$ is deduced by means of the so-called Wagner condition \cite{Wagner,AnnrevKorobkin}, which is nothing but the time integral of the kinematic boundary condition obtained using the velocity field given by equation (\ref{Eq2}), namely,
\begin{equation}
\begin{split}
&\frac{R}{2}\left(\frac{A}{R}\right)^2-VT-\frac{2\,V}{\pi}\left[\int_0^T\frac{\kappa(\tau)\ud\tau}{\sqrt{A(T)^2-\kappa(\tau)^2}} \right.\\
&\left.
-\int_0^T\arcsin\left(\frac{\kappa(\tau)}{A(T)}\right)\ud\tau\right]=0\, ,\label{Eq4}
\end{split}
\end{equation}
with $\kappa(\tau)$ the radius of the wetted area for times $\tau<t$.
Equation (\ref{Eq4}) simply establishes that the wetted radius $A(T)$ is fixed by the instant of time at which a point on the drop interface with initial coordinates $R$=$A(T)$, $Z_d$=$R\,(A(T)/R)^2/2$ reaches $Z$=0.
The dimensionless version of (\ref{Eq4}) then reads,
\begin{equation}
a^2-2t-\frac{4}{\pi}\int_0^a\left[\frac{\kappa}{\sqrt{a^2-\kappa^2}}-\arcsin\left(\frac{\kappa}{a}\right)\right]\frac{\ud\tau}{\ud\kappa}\ud\kappa=0\, ,\label{Eq5}
\end{equation}
where $\ud\tau$=$(\ud\tau/\ud\kappa)\,\ud\kappa$.
The integral equation (\ref{Eq5}), which expresses $t$ as a function of $a$, can be solved by noticing that equation (\ref{Eq5}) possesses a solution of the type $\ud t/\ud a$=$C\,a$. Performing the change of variables $\kappa/a$=$\sin\lambda$, equation (\ref{Eq5}) then reads
\begin{equation}
a^2-2t-\frac{4}{\pi}C\,a^2\int_0^{\pi/2}\left(\sin^2\lambda-\lambda\sin\lambda\cos\lambda\,\right)\ud\lambda=0\,\label{Eq6}
\end{equation}
and, thus, $t$=$a^2\left(1-C/2\right)/2$; therefore, since we assumed that $\ud t/\ud a$=$C\,a$, we conclude that $C$=2/3 and, consequently, $a$=$\sqrt{3\,t}$.

\subsection*{Sheet ejection}

\begin{figure}
\begin{center}
\includegraphics[width=0.48\textwidth]{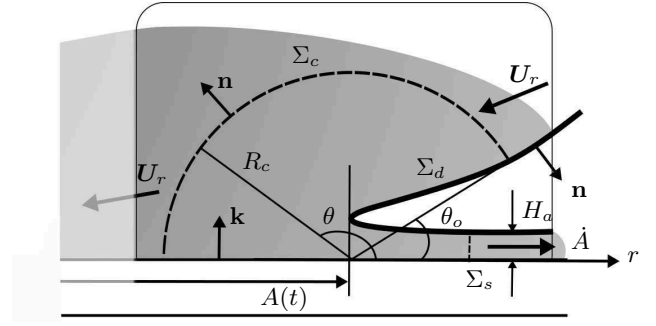}
\caption{\label{figAesq} Definition of the control volume used to determine the thickness $H_a$ of the ejected sheet, which is defined in a frame of reference moving at the velocity $\dot{A}$.\label{Volcontrol}}
\end{center}
\vspace*{-0.7cm}
\end{figure}

The jet ejection process depicted in Fig. 1(c) of the main text, which takes place from $t$=$t_e$ on from the perimeter of the wetted area, namely $r$=$a(t_e)$=$\sqrt{3t_e}$, will be analyzed in a frame of reference moving at a velocity $\ud a/\ud t$. In this frame of reference, the velocity field $\vv{U_r}$ at the surface $\Sigma_c$ sketched in Fig. \ref{Volcontrol} is given by
\begin{equation}
\vv{U_r}=-W\,r^{-1/2}\left(\sin\left(\theta/2\right)\vv{e_r}+\cos\left(\theta/2\right)\vv{e_\theta}\right)-\dot{A}\,\mathbf{e}_x\label{Ur}
\end{equation}
where $r$ is now used to indicate the distance from $r$=$a$ and the expression for $W$=$\sqrt{2a}\,V/\pi$ is obtained taking the limit $r$$\ll$$a$ once the variable $r$ in equation (\ref{Eq2}) is replaced by $r+a$.

It will be shown below that the flow in the control surface $\Sigma$ sketched in Fig. \ref{Volcontrol}, composed by the arc of a circle with radius $r_c$ namely, $\Sigma_c$, the drop surface $\Sigma_d$ as well as by the exit surface $\Sigma_s$, is quasi steady. Thus, neglecting capillarity and viscosity and applying the steady Euler--Bernoulli equation, $P+\rho\,U^2_r/2$=$\rho\,\dot{A}^2$/2
along the constant pressure streamline $\Sigma_d$ with $P$ the liquid pressure, yields that the liquid velocity at the exit surface $\Sigma_s$ in Fig. \ref{figAesq} is uniform and equal to $\dot{A}\,\vv{e_x}$ in the moving frame of reference. Indeed, by virtue of equation (\ref{Ur}), the modulus of the liquid velocity on $\Sigma_d$ can be approximated by $\dot{A}$ in the limit $r_c\gg a^3\propto t^{3/2}$ and, in addition, due to the fact that the streamlines are parallel in the liquid sheet, the pressure drop across the lamella can be neglected. Now note that the integral balances of mass and momentum applied to the control surface $\Sigma$=$\Sigma_c\bigcup\Sigma_d\bigcup\Sigma_s$ sketched in Fig. \ref{Volcontrol}, provide a couple of equations to express $h_{a}$ as a function of $a$. Indeed, using the expression for the relative velocity given in equation (\ref{Ur}), the mass balance yields
\begin{equation}
\begin{split}
\dot{A}\,H_{a}=\int_{\theta_0}^{\pi} \left(Wr_c^{-1/2}\sin\left(\theta/2\right)+\dot{A}\cos\theta\right)\,R_c\,\ud\theta\, ,
\end{split}
\end{equation}
which, in dimensionless form, reads
\begin{equation}
\dot{a}\,h_{a}=2w\,r^{1/2}_c-\dot{a}\,r_c\,\theta_0\, ,\label{Mas1}
\end{equation}
\begin{figure}[t]
\begin{center}
\includegraphics[width=0.48\textwidth]{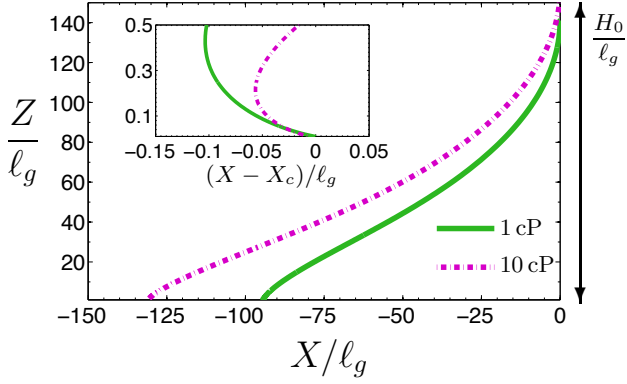}
\caption{\label{figA2} Solution of equation (\ref{thetas}) for a value of the curvature at the tip of the lamella $\ud\gamma/\ud\,s$=-100 and the corresponding critical capillary numbers. In the case of the liquid of viscosity $\mu$=10\,cP, $Ca^*_d$=0.55, and in the case of the liquid with $\mu$=1\,cP, $Ca^*_d$=0.35.}
\end{center}
\end{figure}
with $\theta_0\ll 1$ due to the fact that the drop interface is nearly tangent to the impacting wall [see Fig. \ref{Volcontrol}]. Using the expression for the relative velocity given in equation (\ref{Ur}), as well as the steady Euler--Bernoulli equation $P+\rho\,U^2_r/2$=$\rho\,\dot{A}^2$/2 to calculate pressure on $\Sigma_c$, the momentum balance equation
\begin{equation}
\int_{\theta_0}^\pi\rho\,\vv{e_x}\cdot\vv{U_r}(\vv{U_r}\cdot\mathbf{n})\,R_c\,\ud\theta+\rho\dot{A}^2\,H_{a}=\int_{\theta_0}^\pi\,P\vv{e_x}\cdot\left(-\mathbf{n}\right)\,R_c\,\ud\theta\, ,
\end{equation}
with $\mathbf{e_x}$ a unit vector tangent to the wall yields,
\begin{equation}
-\frac{w^2}{2}\pi+2\,w\dot{a}\,r_c^{1/2}-\dot{a}^2\,r_c\,\theta_0+\dot{a}^2\,h_{a}=0\, . \label{Mom1}
\end{equation}
Substituting the result in equation (\ref{Mas1}) into equation (\ref{Mom1}) leads to the expression for the jet thickness,
\begin{equation}
h_{a}=\frac{\pi\,w^2}{4\,\dot{a}^2}=\frac{a}{2\pi\dot{a}^2}\, . \label{ha}
\end{equation}
Note that, in the deduction above, we have assumed that the flow is quasi steady, and this holds for values of $R_c$ such that
\begin{equation}
\frac{\ud}{\ud T}\int_{\Omega_c}\rho\,\vv{U_r}\,\ud\omega\ll \int_{\Sigma_c}\rho\,\vv{U_r}(\vv{U_r}\cdot\mathbf{n})\ud\sigma\, .
\end{equation}
Since
\begin{equation}
\begin{split}
&\frac{\ud}{\ud T}\int_{\Omega_c}\rho\,\vv{U_r}\,\ud\omega\sim \rho\,V^2\,R \,\ddot{a}\,r_c^{2}\quad \mathrm{and} \\ &\quad \int_{\Sigma_c}\rho\,\vv{U_r}(\vv{U_r}\cdot\mathbf{n})\ud\sigma\sim \rho\,W^2\,R\sim \rho\,V^2\,R\,a\, ,
\end{split}
\end{equation}
the radius $r_c$ thus needs to satisfy the relationship
\begin{equation}
r_c^{2}\ll \frac{a}{\ddot{a}}\rightarrow r_c\ll t\, .
\end{equation}
But $r_c$ has also to satisfy the relationship $r_c\gg h_{a}\sim t^{3/2}$, so it suffices to take $r_c\sim t^{5/4}$.

\subsection*{Dewetting}

\begin{figure}[t]
\begin{center}
\includegraphics[width=0.48\textwidth]{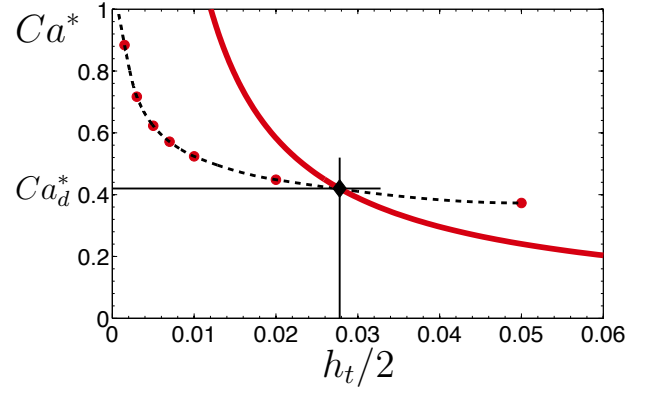}
\caption{Solid points represent the critical capillary number calculated from the numerical solution of equation (\ref{thetas}) for different values of $[\ud\gamma/\ud s(\gamma=\pi/2)]^{-1}$ and $\mu$=5\,cP. The solid curve represents parametrical representation of $\left(\mu\,V/\sigma\right)\sqrt{3}/2\,t_e^{-1/2}$ vs $h_t/2\simeq 1.5\,h_a$=$\sqrt{12}/(2\pi)\,t_e^{3/2}$ (see Fig. 3b of the main text) and the crossing between both functions determine the theoretical capillary number for which the liquid dewets the solid.\label{figA3}}
\end{center}
\end{figure}

Figures 1(e)--(g) in the main text illustrates that the lamella needs to separate from the solid substrate before breaking into drops under the action of capillary forces \cite{Rein}. The dewetting process in our experiments can be qualitatively described using the theory in \cite{Marchand}, which expresses a visco--capillary force balance projected in the direction tangent to the wall. Note that the \emph{tangential} force balance in \cite{Marchand} is different from the \emph{vertical} force balance in the main text.

The equations in \cite{Marchand} permit to determine, in an approximate manner, the critical velocity at which air is entrained when a solid is plunged into a viscous liquid. In this case, the condition for the liquid sheet to detach from the solid substrate is the one determining the air entrainment in an advancing contact line,
\begin{equation}
\frac{\mu\,V_{t}}{\sigma}>Ca^*_{d}\left(q\,\mu_g/\mu,\gamma_0,H_t/\ell_{g,l}\right)\, , \label{condition}
\end{equation}
with $\gamma_0$ the static contact angle and $\ell_{g,l}$ the slip lengths in the solid--gas or solid--liquid interfaces, which are proportional to the mean free path in the case of the gas and, in the case of the liquid, $\ell_l/\ell_{\sigma}\sim 10^{-5}$ \cite{Marchand}. In (\ref{condition}), note also that
\begin{equation}
q=\frac{H}{H+3\ell_g}\, ,
\end{equation}
with $H$ the vertical distance of the free surface to the solid substrate, which is geometrically related to the angle formed by the free interface with the horizontal substrate, $\gamma$, as
\begin{equation}
\frac{\ud h}{\ud s}=\sin\gamma\, .
\end{equation}
The equation for $\gamma(s)$, with $s$ the arclength, is given by the viscocapillary balance projected in the direction tangent to the wall \cite{Marchand}
\begin{equation}
\frac{\ud^2\gamma}{\ud\,s^2}=\frac{3\,Ca}{H\left(H+3\ell_l\right)}f(\gamma,q\,\frac{\mu_g}{\mu})+\cos(\gamma)\,\frac{Bo_t}{h^2_0}\, ,\label{thetas}
\end{equation}
with the function $f$ in (\ref{thetas}) given in \cite{Marchand}, $Bo_t$=$\rho\,|\ud V_t/\ud t|\,H^2_0/\sigma$ and $H_0\simeq H_t/2$ the distance of the tip of the lamella to the wall. The term involving the Bond number is included due to the fact that the flow is described in an accelerated frame of reference, namely, that moving at the velocity of the tip of the lamella with respect to the wall. Since the tip is decelerated by surface tension and $\rho\,|\ud V_t/\ud t|\, H_0\sim \sigma/H_0$, we have limited our computations to $Bo_t$=1.
\begin{table}[t]
\caption{Physical properties of the different gases.}
\begin{ruledtabular}
\begin{center}
\begin{tabular}{cccc}
 &  $\lambda_0$ & $\mu_g$ & $\rho_{g0}$ \\
 & ($\times 10^{-9}$ m) &  ($\times 10^5$ Pa$\cdot$s) &  (kg\,m$^{-3}$) \\
 Helium & 180 & 1.98 & 0.16 \\
 Air & 65 & 1.85 & 1.18 \\
 Krypton  & 55 & 2.51 & 3.42 \\
   SF$_6$ & 39 & 1.53 & 6.04 \\
\end{tabular}
\end{center}
\end{ruledtabular}
The tabulated values correspond to $T_{g0}=298.15$ K, $p_{g0}=10^{5}$ Pa. Therefore, for different values of the gas temperature $T_g$ and pressure $p_g$, $\lambda=\lambda_0 \left(T_g/T_{g0}\right)\left(p_{g0}/p_g\right)$ and $\rho_g=\rho_{g0}\left(T_{g0}/T_{g}\right)\left(p_{g}/p_{g0}\right)$.
\label{tab2}
\end{table}

The main difference between our calculations and those in \cite{Marchand} are the boundary conditions that need to be satisfied by equation (\ref{thetas}). Indeed, in our case, equation (\ref{thetas}) has been solved fixing $Ca$ and shooting from a position at the interface with $\gamma$=$\pi/2$, $\ud \gamma/\ud s$=$-2/h_t$ and varying the height $h_0(\gamma=\pi/2)$ until $\gamma$=$\gamma_0$ at the wall, with the static contact angle fixed to $\gamma_0$=$\pi/6$ in all the calculations presented in this section. The value of the capillary number is increased until the system (\ref{condition})--(\ref{thetas}) fails to converge to the fixed value of $\gamma_0$, a condition that determines the value of the critical capillary number $Ca^*_d$. Figure \ref{figA2} depicts the computed local shapes of the tip of the lamella at the corresponding critical capillary numbers for the same value of the initial curvature, $|\ud\gamma/\ud\,s|$=100 and two liquid viscosities, $\mu$=10\,cP and $\mu$=1\,cP, showing that the angle of the wedge formed between the edge of the lamella and the solid substrate varies only slightly with the viscosity ratio. Figure \ref{figA3} illustrates the critical capillary number calculated solving equation (\ref{thetas}) for $\mu$=5\,cP and different values of the interfacial curvature $\ud\gamma/\ud s(\gamma=\pi/2)$. Figure \ref{figA3} also shows that the critical capillary number increases with $\ud\gamma/\ud s(\gamma=\pi/2)$. To determine the approximate experimental value of $Ca^*_d$ for the case of silicon oil drops with $\mu$=4.6\,cP, the values of $\left(\mu V/\sigma\right) \sqrt{3}/2 t_e^{-1/2}(Re,We)$ vs $h_t/2$, with $h_t$=$3\,h_a$=$\sqrt{12}/\pi\,t_e^{3/2}(Re,We)$ and $t_e$ given by equation (3) of the main text are represented in the same Fig. \ref{figA3}. The crossing between the two curves fixes the critical capillary number $Ca^*_d$$\simeq$0.42, which is far smaller than the critical capillary number at which splash is experimentally observed (see the inset in Fig. 4(a) of the main text). These results reinforce our starting hypothesis that the liquid dewets the substrate for values of the capillary number smaller than those for which the splash transition is experimentally observed.

\subsection*{Determination of $K_u$ and $K_l$}

\begin{figure}[t!]
\begin{center}
\includegraphics[width=0.46\textwidth]{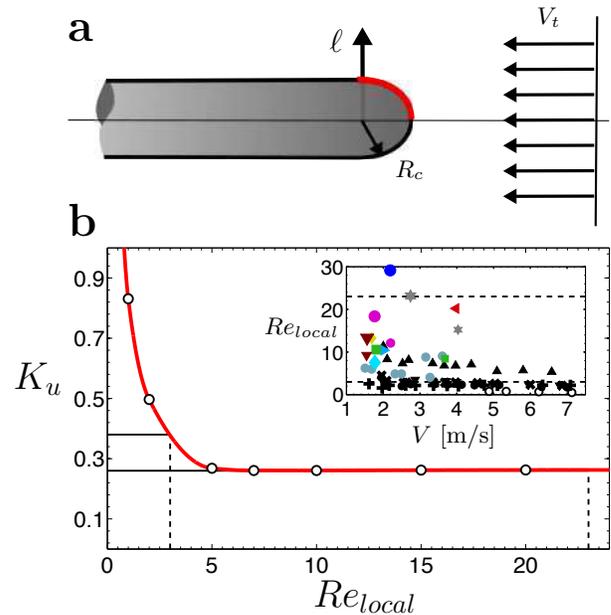}
\caption{\label{figmissile} (a) Figure sketching the top part of the advancing lamella. (b) Values of the local Reynolds number for the experimental data considered in this study.}
\end{center}
\end{figure}

\begin{figure}[t]
\begin{center}
\includegraphics[width=0.48\textwidth]{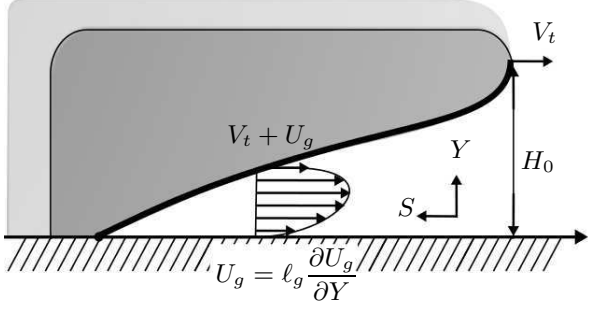}
\caption{\label{figAesq_bis} Figure sketching the flow in the lubricating air layer located between the substrate and the lamella.}
\end{center}
\end{figure}

The lift forces exerted by the relative gas flow in the regions sketched in Figs. \ref{figmissile}--\ref{figAesq_bis}, will be calculated next. The force in the quarter of a circle marked in red in Fig. \ref{figmissile} has been computed numerically, using the comercial code Fluent, for several values of the gas Reynolds number $Re_{local}=\rho_g V_t\,R_c/\mu_g$, with $R_c\simeq H_t$ the radius of curvature of the front part of the advancing lamella. The result, depicted in Fig.  (\ref{figmissile})b, reveals that $K_u\simeq 0.3$ for the range of local Reynolds numbers relevant to drop splashing, namely, $3<Re_{local}<10^2$.

To calculate the lift force in the region located between the lamella and the substrate, the shape of the advancing front is approximated to that of a wedge of angle $\alpha$ such that $\tan(\alpha)=H_0/L$, with $H_0\propto H_t$ (see Fig. \ref{figAesq_bis}). Under the lubrication approximation, the general form of the velocity field,
\begin{equation}
U_g(Y,S)=U_0(S)+U_1(S)\,Y+\frac{Y^2}{2\mu_g}\frac{\partial\,P_g}{\partial\,S}\, ,\label{Ug}
\end{equation}
needs to satisfy the boundary conditions at $Y$=0 and $Y$=$H(S)$=$H_0\left(1-S/L\right)$,
\begin{equation}
\begin{split}
&Y=0\, , \quad U_g=\ell_g\frac{\partial U_g}{\partial\,Y}\quad\mathrm{and}\\ & Y=H(S)\, ,\quad U_g=-V_t-\ell_\mu\frac{\partial U_g}{\partial\,Y}\, .\label{BCUg}
\end{split}
\end{equation}
with $\ell_g\simeq 1.2 \lambda$ the slip length of the gas\cite{Maurer}, $\lambda=k_B\,T_g/(\sqrt{2}\pi d^2\,p_g)$ the mean free path between gas molecules, $k_B$ Boltzmann constant, $T_g$ and $p_g$ the gas temperature and pressure respectively, $d$ the effective diameter of gas molecules (see Table \ref{tab2}) and $\ell_\mu$ arising from the continuity of shear stresses at the interface, which demands that, at $Y$=$H(S)$,
\begin{equation}
\mu_g\frac{\partial U_g}{\partial Y}\simeq \mu\frac{\partial V}{\partial Y}\rightarrow V_s\sim \frac{\mu_g}{\mu}\,H_t\,\frac{\partial U_g}{\partial Y}\, .\label{shearstress}
\end{equation}
In (\ref{shearstress}), it has been assumed that $\partial V/\partial Y\sim V_s/H_t$, with $V_s$ the liquid velocity at the interface in a frame of reference moving at $V_t$ and, thus, $\ell_\mu\sim H_t\,\mu_g/\mu$. Therefore, the gas velocity field that satisfies the boundary conditions given in (\ref{BCUg}) expressed in a frame of reference moving at $V_t$ is
\begin{equation}
\begin{split}
& U_g=-V_t\,\frac{Y-\ell_\mu-H}{H+\ell_g+\ell_\mu}-\\&-\left(\frac{H}{\mu_g}\right)\frac{Y+\ell_g}{H+\ell_g+\ell_\mu}\left(\ell_\mu+\frac{H}{2}\right)\frac{\partial\,P_g}{\partial\,S}+\frac{Y^2}{2\,\mu_g}\frac{\partial\,P_g}{\partial\,S}\, .\label{Ugas}
\end{split}
\end{equation}
The pressure gradient $\partial\,P_g/\partial\,S$ can be deduced imposing that the net flow rate per unit length is zero in the moving frame of reference, namely,
\begin{equation}
q=\int_0^H U_g(S,Y)\,\ud Y=0\, ,
\end{equation}
with $U_g$ given by equation (\ref{Ugas}) from which we obtain
\begin{equation}
\frac{\ud\bar{\pi}_g}{\ud\,\bar{s}}=\frac{\bar{h}+2\,\bar{\ell}_\mu}{\bar{h}\left(\bar{h}^2+4\,\bar{h}(\bar{\ell}_\mu+\bar{\ell}_g)+12\,\bar{\ell}_g\bar{\ell}_\mu\right)}\, ,\label{pis}
\end{equation}
where
\begin{equation}
\begin{split}
&P_g=\frac{6\,V_t\,\mu_g}{\tan^2(\alpha)}\bar{\pi}_g\, ,\quad S=L\,\bar{s}\, ,\quad H=H_0\,\bar{h}\quad\,\\&\ell_\mu=H_0\bar{\ell}_\mu\quad \mathrm{and}\quad \ell_g=H_0\,\bar{\ell}_g\, .
\end{split}
\end{equation}

Thus, the vertical force per unit length exerted by the pressure distribution obtained from the integration of equation (\ref{pis}), is approximately given by
\begin{equation}
F_v=\frac{6\,\mu_g\,V_t}{\tan^2(\alpha)}\int_0^1\,\bar{\pi}_g\,\ud\bar{s}=\frac{\mu_g\,V_t}{\tan^2(\alpha)}\, K_l\, ,
\end{equation}
where
\begin{equation}
\begin{split}
&K_l=-\left(6/\tan^2(\alpha)\right)\times\\&\times\left(C_2\left[a\ln(1+a)-a\ln a\right]+C_3\left[b\ln(1+b)-b\ln b\right]\right)\, ,
\end{split}
\end{equation}
with
\begin{equation}
\begin{split}
&a=\left(\bar{\ell}_g+\bar{\ell}_\mu\right)+2\sqrt{\left(\bar{\ell}_g-\bar{\ell}_\mu\right)^2+\bar{\ell}_g\bar{\ell}_\mu}\, \\ & b=\left(\bar{\ell}_g+\bar{\ell}_\mu\right)-2\sqrt{\left(\bar{\ell}_g-\bar{\ell}_\mu\right)^2+\bar{\ell}_g\bar{\ell}_\mu}\, ,
\end{split}
\end{equation}
and
\begin{equation}
C_1=\frac{2\,\bar{\ell}_\mu}{a\,b}\, ,\quad C_2=\frac{1-C_1}{b-a}\, ,\quad C_3=-\left(C_1+C_2\right)\, .
\end{equation}
To simplify the expression for $K_l$, we set $\bar{\ell}_\mu=0$ in equation (\ref{pis}), yielding, for $H_0=H_t/4$
\begin{equation}
K_l=-(6/\tan^2(\alpha))\,\left(\ln\left[16\ell_g/H_t\right]-\ln\left[1+16\ell_g/H_t\right]\right)\, .\label{Kl}
\end{equation}
The theoretical results presented in Fig. 4 of the main text have been obtained using equation (\ref{Kl}) with $\alpha=60^o$, independent on the type of fluid considered (as suggested by the results in Fig. \ref{figA2}) and for a value of $K_u=0.3$ (see figure \ref{figmissile}). The angle $\alpha$ should be dependent on the wetting properties of the solid but, in the case of our experiments, all the liquids partially wet the substrate with a similar contact angle $\sim 20^o$.

To conclude, note that corrections to the mean free path as a consequence of the modification of both pressure and temperature along the coordinate $\bar{s}$ could have also been included in the formulation\cite{Maurer}, but in order to deduce an analytic expression for $K_l$ and since the improvement in the results should not be significative, we only report here the result corresponding to $\ell_g$ independent on $\bar{s}$.\\

\subsection*{Comparison of experimental data with existing correlations and with the theory in the main text}

\begin{figure}[t]
\begin{center}
\includegraphics[width=0.48\textwidth]{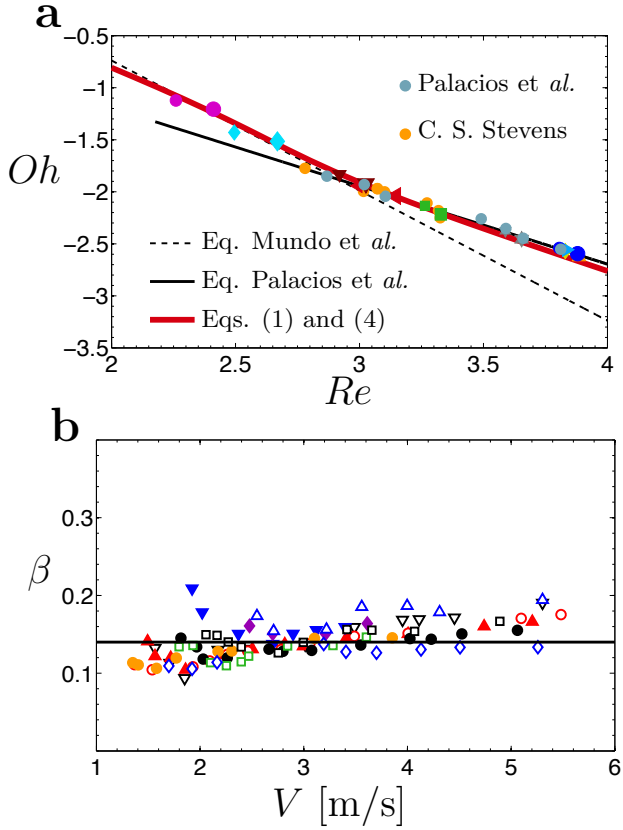}
\caption{\label{fig11} (a) Comparison in the $Oh$--$Re$ plane between the experimental critical velocity and that calculated (in red) using Eqs. (1) and (4) in the main text. The correlations $57.7\,Re^{-1.25}$ by Mundo \emph{et al.} \cite{Mundo} and $3.39\,Re^{-0.75}$ by Palacios \emph{et al.} \cite{Palacios} are also represented. (b) Values of $\beta$ calculated through Eqs. (1) and (4) in the main text using the experimental data by Stevens \cite{Stevens}. Observe that $\beta\simeq 0.14$ for all the experimental data.}
\end{center}
\end{figure}

Figure \ref{fig11}a compares, in the $Re$--$Oh$ plane, the splash threshold predicted by the correlations in \cite{Mundo} and \cite{Palacios} with our own data and also with the experiments in \cite{Palacios} and in \cite{Stevens} for air at normal conditions. It is clearly observed that each of the correlations follow, in a different range of $Oh$, the experiments. Our theory, in red, predicts the experimental observations and reproduce both correlations in the whole range of $Oh$ investigated. Our calculation is the result of solving equations (1) and (4) in the main text for air properties at normal atmospheric conditions using the material properties of ethanol. Also note from figure \ref{fig11}b that the splash threshold corresponding to the experimental data in Stevens\cite{Stevens}, where the critical speed of drops of different liquids falling within several gases and different pressures is investigated, is also characterized by $\beta\simeq 0.14$.

\begin{figure}[t]
\begin{center}
\includegraphics[width=0.48\textwidth]{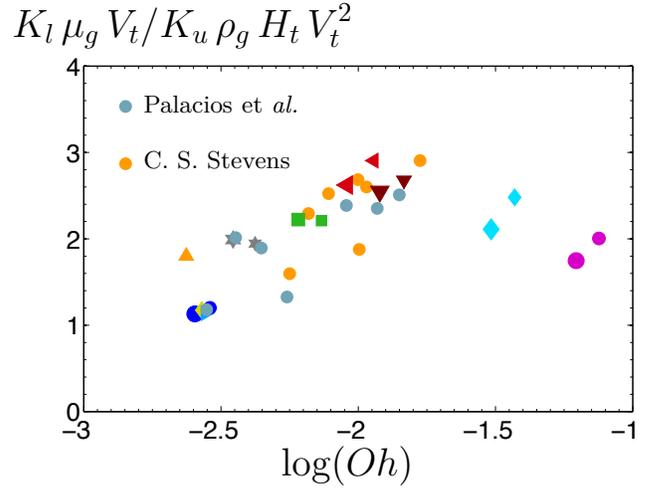}
\caption{\label{KuKl} Ratio $K_l\,u_g\,V_t/\left(K_u\,\rho_g\,V^2_t\,H_t\right)$ vs $Oh$.}
\end{center}
\end{figure}

As pointed out in the main text, due to the fact that splashing is triggered by the relative motion between the liquid and the outer gaseous atmosphere, the splash criterion cannot be expressed as $Oh\,Re^{\kappa}=K$, with $\kappa$ an arbitrary exponent and $K$ a constant, since this would imply that the critical velocity depends only on the liquid properties and $R$. With the purpose of expressing our splash criterion in a more compact, albeit only approximate form, Fig. \ref{KuKl} shows that the contribution to the lift force associated to the gas flow in the wedge is larger than the force imparted by the gas on the top part of edge of the liquid sheet, with the only exception of water (the liquid with the largest value of $\sigma$ considered in this study and the one with the largest critical speed). A plausible approximation in view of Fig. \ref{KuKl} would be to express the total lift as $\mathbf{\ell}\propto K_l\,\mu_g\,V_t$ which, using the low--$Oh$ and high--$Oh$ limits for the ejection time deduced from Eq. (1) in the main text, and taking into account the logarithmic dependence of $K_l$, yields the following \emph{approximate} splash criteria:
\begin{equation}
\begin{split}
& \mathrm{High}\quad Oh:\quad \mu/\mu_g\propto \left(Re\,Oh_g^{8/5}\right)^{-5/4}\\
& \mathrm{Low}\quad Oh:\quad \mu/\mu_g\propto \left(Re\,Oh_g^{8/5}\right)^{-1}\, , \label{limits}
\end{split}
\end{equation}
with $Oh_g=\mu_g/\sqrt{\rho R\sigma}$.

\begin{figure}[t]
\begin{center}
\includegraphics[width=0.48\textwidth]{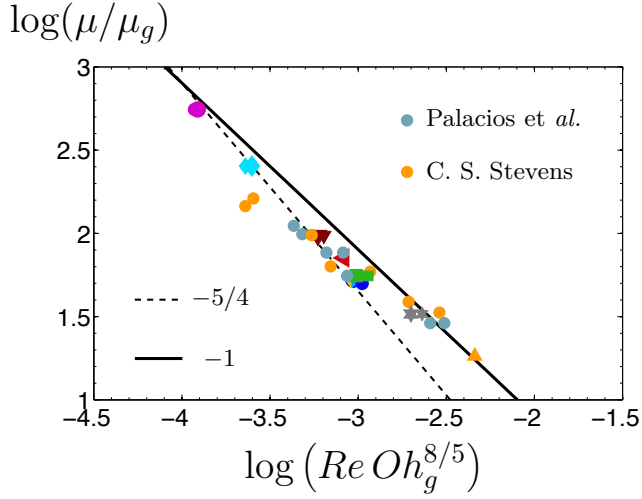}
\caption{\label{Ohg} Comparison of the experimental splash threshold with the approximate expressions deduced in the main text and also reproduced in equation (\ref{limits}).}
\end{center}
\end{figure}

The agreement between the experimental data and the predictions in equation (\ref{limits}) is reasonably good, with the exponent affecting the modified Reynolds number $Re\,Oh_g^{8/5}$ in the high--$Oh$ limit, identical to that in the correlation by \cite{Mundo}. This means that if the experimental data was such that $\sqrt{\rho\,R\,\sigma}\sim$ $constant$, our result would reproduce the scaling in \cite{Mundo}. However, it needs to be pointed out that the contribution to the total lift force associated to the suction at the top part of the lamella cannot be neglected. Indeed, as it is shown in figure 7(a) as well as in the main text, the agreement between theory and experiments is even better than the one depicted in figure (\ref{Ohg}) when the full expression for the lift force given by Eq. (2) in the main text is used instead of $\mathbf{\ell}\propto K_l\,\mu_g\,V_t$.

\clearpage
\newpage

\bibliography{main}

\end{document}